\documentclass{egpubl}
\usepackage{pg2025}

\ConferencePaper        % uncomment for (final) Conference Paper
\SpecialIssuePaper         % uncomment for final version of Computer Graphics Forum, special issue
\CGFccby
\usepackage[T1]{fontenc}
\usepackage{dfadobe}  

\usepackage{cite}  % comment out for biblatex with backend=biber
\BibtexOrBiblatex
\electronicVersion
\PrintedOrElectronic
\ifpdf \usepackage[pdftex]{graphicx} \pdfcompresslevel=9
\else \usepackage[dvips]{graphicx} \fi

\usepackage{egweblnk} 
\title{Improved 3D Scene Stylization via Text-Guided Generative Image Editing with Region-Based Control}

\author[H. Fujiwara et al.]
 {\parbox{\textwidth}{\centering Haruo Fujiwara$^{1}$\orcid{0009-0001-0443-4477},
Yusuke Mukuta$^{1,2}$\orcid{0000-0002-7727-5681}, and
Tatsuya Harada$^{1,2}$\orcid{0000-0002-3712-3691}
         }
         \\
 {\parbox{\textwidth}{\centering $^1$ The University of Tokyo, Japan\\
          $^2$ RIKEN AIP, Japan
        }
 }
}
\usepackage{amsmath}
\usepackage{amssymb}
\usepackage{amsfonts}
\usepackage{makecell}  % For multi-line text in table cells
\usepackage{array}     % Optional: for better column formatting
\usepackage[linesnumbered,ruled,vlined]{algorithm2e}

\begin{document}

 %uncomment for using teaser
 \teaser{
  \includegraphics[width=0.9\linewidth]{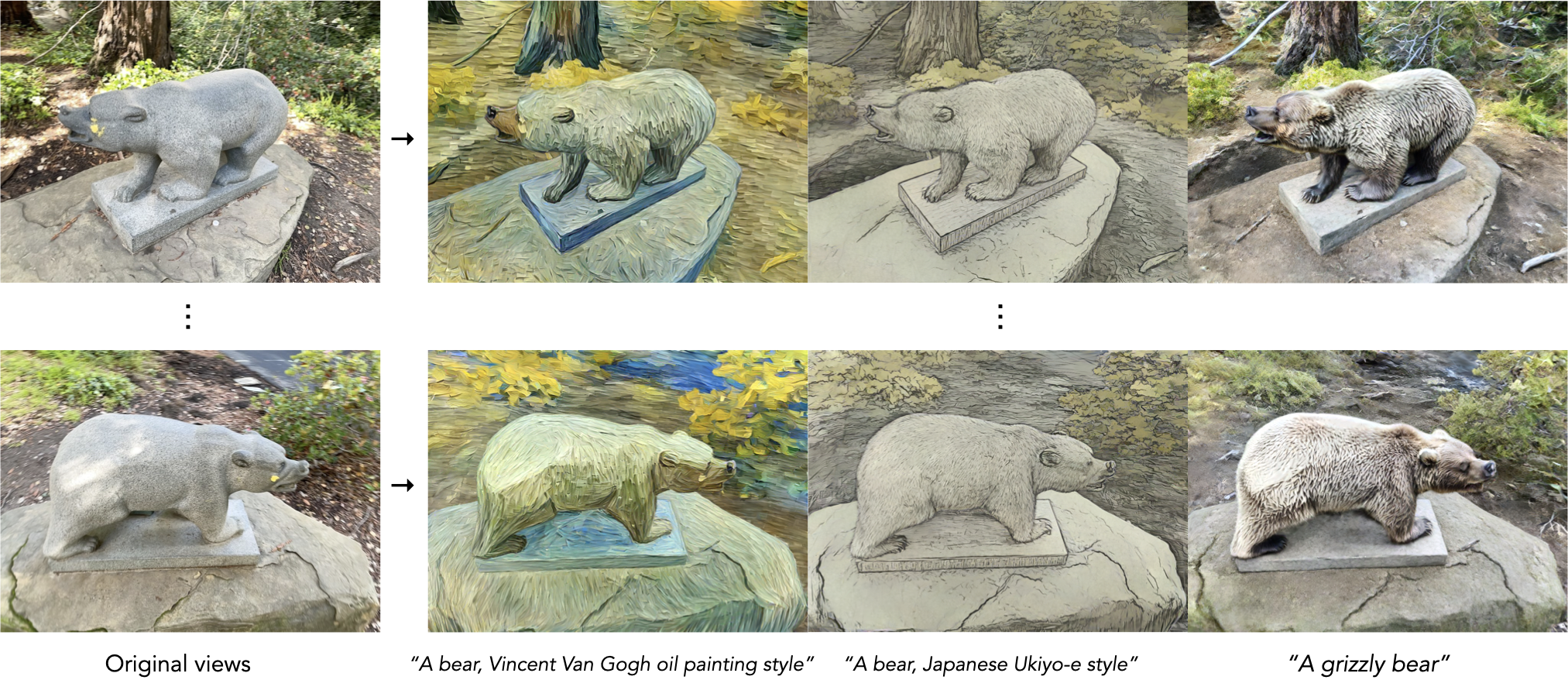}
  \centering
   \caption{We present a robust text-driven 3D scene stylization technique based on the Gaussian splatting representation. Our method enables high-quality, semantically consistent stylization of 3D scenes guided by text prompts, while preserving view consistency and geometric fidelity.}
 \label{fig:teaser}
}

\maketitle
%-------------------------------------------------------------------------
\begin{abstract}
Recent advances in text-driven 3D scene editing and stylization, which leverage the powerful capabilities of 2D generative models, have demonstrated promising outcomes. However, challenges remain in ensuring high-quality stylization and view consistency simultaneously. Moreover, applying style consistently to different regions or objects in the scene with semantic correspondence is a challenging task. To address these limitations, we introduce techniques that enhance the quality of 3D stylization while maintaining view consistency and providing optional region-controlled style transfer. Our method achieves stylization by re-training an initial 3D representation using stylized multi-view 2D images of the source views. Therefore, ensuring both style consistency and view consistency of stylized multi-view images is crucial. We achieve this by extending the style-aligned depth-conditioned view generation framework, replacing the fully shared attention mechanism with a single reference-based attention-sharing mechanism, which effectively aligns style across different viewpoints. Additionally, inspired by recent 3D inpainting methods, we utilize a grid of multiple depth maps as a single-image reference to further strengthen view consistency among stylized images. Finally, we propose Multi-Region Importance-Weighted Sliced Wasserstein Distance Loss, allowing styles to be applied to distinct image regions using segmentation masks from off-the-shelf models. We demonstrate that this optional feature enhances the faithfulness of style transfer and enables the mixing of different styles across distinct regions of the scene. Experimental evaluations, both qualitative and quantitative, demonstrate that our pipeline effectively improves the results of text-driven 3D stylization.
\noindent\textbf{\href{https://haruolabs.github.io/improved-gs-style-page/}{Project page}}
%and faithfully replicates the distinctive styles of historical artists.
%-------------------------------------------------------------------------
%  ACM CCS 1998
%  (see https://www.acm.org/publications/computing-classification-system/1998)
% \begin{classification} % according to https://www.acm.org/publications/computing-classification-system/1998
% \CCScat{Computer Graphics}{I.3.3}{Picture/Image Generation}{Line and curve generation}
% \end{classification}
%-------------------------------------------------------------------------
%  ACM CCS 2012
%   (see https://www.acm.org/publications/class-2012)
%The tool at \url{http://dl.acm.org/ccs.cfm} can be used to generate
% CCS codes.
%Example:


\end{abstract}  
%-------------------------------------------------------------------------
% Content

\section{Introduction}
Recent 3D reconstruction techniques, such as Neural Radiance Field (NeRF) \cite{mildenhall2020nerf} and Gaussian splatting \cite{kerbl3Dgaussians}, enable the creation of high-quality digital representations of real scenes from 2D photos.
Editing and stylizing such 3D representations have emerged as an important research focus for applications such as content creation and 3D post-editing.
However, direct manipulation of 3D representations is not straightforward as multi-view consistency of both geometry and color need to considered.
To alleviate such a bottleneck and bypass the necessity for scarce 3D data, recent efforts often utilize foundational 2D generative models such as Stable Diffusion to achieve 3D editing and stylization via knowledge distillation.
While these methods have pioneered 3D editing, most techniques rely on iterative techniques such as SDS \cite{poole2022dreamfusion} and iterative dataset updates \cite{instructnerf2023, igs2gs} which often require significant time until convergence.
As an alternative, some techniques \cite{fujiwara2024sn2n, chen2024dge} propose a more simplified sequential process: (a) consistent multi-view edit, and then (b) 3D re-train on edited images.
This approach benefits users by providing a preview of the stylized scene before full 3D re-training.

In our work, we also follow a generate-then-train approach similar to Style-NeRF2NeRF \cite{fujiwara2024sn2n} where we first stylize 2D multi-view images and then fine-tune the source Gaussian Splatting \cite{kerbl3Dgaussians, huang20242d} scene based on these images.
We observe that consistency of stylized multi-view images contributes to effective 3D stylization.
To this end, we propose a training-free technique that uses a text-driven generation pipeline conditioned on tiled depth maps for consistent view stylization. %(By incorporating an attention-sharing mechanism, our method improves style consistency across generated views.)

Along with the above improvements, we propose a multi-region importance-weighted sliced Wasserstein distance loss (MR-IW-SWD) to perform semantically consistent style transfer.
While prior work \cite{fujiwara2024sn2n} effectively applies SWD for 3D style transfer, it lacks region-specific control—a key limitation in NeRF-based representations, where stylization is applied uniformly across sampled patches.
To address this, we extend the SWD loss with segmentation masks for semantically aware, region-specific style transfer.
Additionally, we adopt importance-weighted sampling to improve efficiency by prioritizing perceptually salient projection directions. These enhancements, integrated into our MR-IW-SWD, enable efficient and semantically consistent 3D stylization.
Our contributions are summarized as following:

\begin{itemize}

 \item We present an improved training-free style-aligned diffusion pipeline for editing source views with improved multi-view style consistency.

 \item We propose multi-region importance-weighted sliced Wasserstein distance loss for additional semantic consistency and spatial control during 3D style transfer with enhanced training efficiency.

 \item We show that Gaussian Splatting scenes finetuned with stylized views using our 2D generation pipeline shows competitive 3D style transfer performance.

\end{itemize}

\section{Related Work}
\subsection{2D Text-to-Image Generation}
Diffusion models \cite{sohl2015deep, song2020score, dhariwal2021diffusion} are powerful generative models recognized for producing high-quality, diverse images. Drawing on concepts from non-equilibrium thermodynamics, these models generate images by reversing a diffusion process, gradually denoising from random noise to a coherent image. Typically, they employ classifier-free guidance \cite{ho2022classifier} for text-conditioned image generation. Extensions like ControlNet \cite{zhang2023adding} allow for more detailed control by incorporating additional conditions, such as depth maps, normal maps, and feature lines. Similarly, IP-Adapter \cite{ye2023ip-adapter} enables enhanced control by using an image as a prompt.

\subsection{Style Transfer}
Style transfer blends a source image with a style image, creating a result that retains the content of the source but reflects the style of the other. Since the foundational neural algorithm by Gatys et al. \cite{gatys2015neural}, research has focused on enhancing 2D style transfer including faster optimization \cite{johnson2016perceptual_faststyle, huang2017arbitrary_adain}, zero-shot transfer \cite{li2017universal}, and photorealism \cite{luan2017deepphoto}.
While most previous methods leveraged the feature distributions of a pretrained convolutional model such as VGG19 \cite{simonyan2014very} to manipulate style, recent advances in diffusion models have introduced enhanced control over stylized image generation using both text and reference images. Lightweight fine-tuning methods \cite{frenkel2024implicit, shah2023ziplora, Everaert_2023_ICCV, sohn2024styledrop, kawar2023imagic} adapt the underlying diffusion models for personalized stylization. Pre-trained add-ons \cite{xing2024csgo, ye2023ip-adapter} allow flexible style manipulation by referencing text or an image directly. Training-free approaches based on latent code inversion \cite{mokady2023null, cheng2023general_i2i, tumanyan2023plug, parmar2023zero_i2i} also facilitate style control, while others \cite{cao2023masactrl, hertz2024style, hertz2022prompt, hertz2023delta, meng2021sdedit} use attention modulation for controlled generation without finetuning. Our work leverages a training-free approach with attention-sharing \cite{hertz2024style} to generate consistently stylized images across multi-views.

\subsection{3D Representation}
Modern 3D representations such as Neural Radiance Fields (NeRF) \cite{mildenhall2020nerf} and 3D Gaussian Splatting \cite{kerbl3Dgaussians} have become increasingly popular due to their advantages over traditional approaches like polygon meshes or voxel grids. NeRF, as an implicit representation, encodes scenes in a continuous volumetric function, enabling photo-realistic novel view synthesis at arbitrary resolutions with a compact memory footprint. More recently, 3D Gaussian Splatting has emerged as a powerful explicit alternative that supports real-time, high-quality rendering by leveraging conventional graphics pipelines. In this work, we adopt 2D Gaussian Splatting \cite{huang20242d} as our base representation—a recent variant that achieves superior geometry reconstruction while maintaining competitive performance and view synthesis quality.

\subsection{3D Stylization and Editing}
Recent research has extended 2D style transfer techniques to 3D by embedding deep feature statistics into volumetric scene representations, particularly with NeRF-based models \cite{liu2023stylerf, wang2023nerfart, zhang2022arf, chiang2022stylizing, huang2022stylizednerf, nguyen2022snerf, pang2023locally, kim2024fprf, zhang2023ref} and more recently with Gaussian Splatting \cite{gu2024artnvg,chen2024gaussianeditor,wang2024gaussianeditor,chen2024dge, wang2024view, liu2024stylegaussian, galerne2025sgsst, kovacs2024��}. Beyond reference image-based stylization, text-driven 3D editing methods have emerged, leveraging the generative power of 2D text-to-image (T2I) models. Some methods \cite{instructnerf2023, dong2024vica} alternate between scene finetuning and multi-view supervision, while others such as \cite{Koo:2024PDS} directly update the underlying NeRF representation in a single optimization loop. In contrast, generate-then-train approaches \cite{signerf, fujiwara2024sn2n, chen2024dge, wynn2025morpheus} decouple the process by first generating stylized views and then refining the 3D scene. This strategy offers a more intuitive and interactive workflow by allowing users to preview stylization effects before full 3D refinement. In our work, we adopt this sequential paradigm and propose a training-free diffusion pipeline for generating style-consistent multi-view images. These images are then used to fine-tune the 3D scene using our improved style transfer loss, resulting in high-quality and semantically coherent 3D stylization.

\section{Method}
\subsection{Preliminaries}
\subsubsection{Gaussian Splatting}
3D Gaussian Splatting (3DGS) represents a scene using a set of anisotropic 3D Gaussians, which are typically initialized from a sparse point cloud reconstructed via Structure-from-Motion (SfM) \cite{schoenberger2016sfm}.
Each Gaussian is explicitly parameterized by its 3D position $\mathbf{x}_k$ and a covariance matrix $\Sigma$ defined as 
$\Sigma = \mathbf{R} \mathbf{S} \mathbf{S}^\top \mathbf{R}^\top$:

\begin{equation}
    g_k(\mathbf{x}) = \exp{\big( -\frac{1}{2}(\mathbf{x}-\mathbf{x}_k)^\top  \Sigma^{-1} (\mathbf{x} - \mathbf{x}_k) \big)}
\end{equation}
where $\mathbf{S}$ is a scaling matrix and $\mathbf{R}$ is a rotation matrix.
To render an image, the Gaussians are transformed into camera coordinates using the world-to-camera transformation matrix $\mathbf{W}$ and projected onto the image plane via a local affine projection $\mathbf{J}$ \cite{zwicker2001ewa}, yielding a transformed covariance 
$\Sigma' = \mathbf{J} \mathbf{W} \Sigma \mathbf{W}^\top \mathbf{J}^\top$.
In this work, we adopt 2D Gaussian Splatting (2DGS) \cite{huang2017arbitrary_adain} as our base representation, which is a variant that offers view-consistent geometry. Specifically, 2DGS omits the third row and column of $\Sigma'$, and represents the projected Gaussians $g^{2D}$ using a 2D covariance matrix $\Sigma^{2D}$.
Rendering is achieved via volumetric alpha blending, integrating view-dependent appearance values $\mathbf{c}_k$ from front to back, weighted by their corresponding alpha values $\alpha_k$, as follows:

\begin{equation}
    c(\mathbf{x}) = \sum_{k=1}^K \mathbf{c}_k \alpha_k g_k^{2D}(\mathbf{x}) \prod_{j=1}^{k-1}(1-\alpha_j g_j^{2D}(\mathbf{x}))
\label{eq:volrender}
\end{equation}

We obtain the source 3D scene by optimizing the 2DGS representation using the loss functions proposed in the original work, which guide the reconstruction toward photorealistic and geometrically consistent results.

\subsubsection{Multi-View Stylization}
Recent works have explored leveraging powerful 2D generative models, such as diffusion models, for text-driven 3D scene editing and stylization. While these methods show promising results, challenges persist in ensuring high-quality stylization, maintaining view consistency, and applying style consistently across different regions or objects within a scene. Style-NeRF2NeRF \cite{fujiwara2024sn2n} proposes an effective pipeline to address these challenges by stylizing a 3D scene using 2D image diffusion models \cite{podell2023sdxl}. The pipeline begins with a Neural Radiance Field (NeRF) model reconstructed from a set of multi-view images. Style transfer is then achieved by refining the source NeRF model using stylized images generated by a style-aligned image-to-image diffusion model. A key step involves generating perceptually similar multi-view images based on a given target style prompt. This is accomplished by employing a depth-conditioned diffusion model enhanced with an attention-sharing mechanism. Following the generation of stylized multi-view images, the style transfer process is guided using the sliced Wasserstein loss, which is computed based on feature maps extracted from a pre-trained Convolutional Neural Network (CNN) model \cite{simonyan2014very}. A notable aspect of Style-NeRF2NeRF pipeline is its decoupled nature, allowing users the flexibility to experiment with various style prompts and preview the stylized 3D results before proceeding to the NeRF fine-tuning stage. This method has demonstrated the ability to transfer diverse artistic styles to real-world 3D scenes while achieving competitive quality.

\subsubsection{Sliced Wasserstein Distance-Based Style Transfer}
Feature statistics from pre-trained Convolutional Neural Networks (CNNs), such as VGG-19 \cite{simonyan2014very}, are well-known for capturing the style of an image \cite{gatys2015neural,johnson2016perceptual_faststyle,huang2017arbitrary_adain,li2017universal,luan2017deepphoto}.
Sliced Wasserstein Distance (SWD) loss has been proposed as an effective alternative for learning image style, especially in applications such as texture synthesis \cite{heitz2021sliced} and 3D style transfer \cite{fujiwara2024sn2n}. In our work, we build upon SWD to lift the stylized 2D multi-view images into a 3D representation with additional spatial control.

Let $F^l_m \in \mathbb{R}^{N_l} (m=1,\ldots,M_l)$ represent the feature vector at pixel $m$ in the $l$-th layer, where $M_l$ is the number of pixels and $N_l$ is the feature dimension. The discrete probability density function $p^l(x)$ for these features in layer $l$ can be formulated with the Dirac delta function as:

\begin{equation}
    p^l(x) = \frac{1}{M_l} \sum_{m=1}^{M_l} \delta_{F^l_m}(x)
\label{eq:feature-stats}
\end{equation}

Using the feature distributions $p^l, \hat{p}^l$ from images $I$ and $\hat{I}$, the style loss is defined as the sum of SWD over layers:

\begin{equation}
    \mathcal{L}_{style} = \sum_{l=1}^{L} \mathcal{L}_{SWD}(p^l, \hat{p}^l)
\end{equation}
where $\mathcal{L}_{SWD}$ measures 1D Wasserstein distances between projected features, calculated over random directions $V \in \mathbb{S}^{N_l-1}$ uniformly sampled from a unit hypersphere.
Using scalar features $p^l_V = \{ \langle F^l_m, V \rangle \} \in \mathbb{R}^{M_l}$ projected by $V$, $\mathcal{L}_{SWD}$ can be written as:

\begin{equation}
    \mathcal{L}_{SWD} = \sum_{l=1}^{L} \mathbb{E}_V[ \mathcal{L}_{SW1D}(p^l_V, \hat{p}^l_V) ]
\label{eq:sw}
\end{equation}
where the 1D Wasserstein distance $\mathcal{L}_{SW1D}$ is simply the $L^2$ distance of vectors with sorted scalar values from $p^l_V$ and $\hat{p}^l_V$:

\begin{equation}
    \mathcal{L}_{SW1D}(p^l_V, \hat{p}^l_V) = \frac{1}{|p^l_V|} \lVert \texttt{sort}(p^l_V) - \texttt{sort}(\hat{p}^l_V) \rVert^2
\label{eq:sw1d}
\end{equation}

Taking the expectation over projections $V$ approximates the target distribution effectively in practice. Unlike the Gram matrix, SWD converges to the complete target distribution and has computational efficiency, scaling at $\mathcal{O}(M \log M)$ for an $M$-dimensional distribution, which suits gradient-based learning. 

\subsection{Multi-View Editing With Tiled Depth Reference}
Modern image generation models \cite{rombach2022high, podell2023sdxl, esser2024scaling} exhibit a degree of 3D awareness, attributed to training on large-scale image-text datasets \cite{zhan2023does, zhan2023general}.
Conditioning these models with 3D cues such as depth maps via ControlNet \cite{zhang2023adding} has been shown to significantly improve multi-view consistency.
Building on this capability, several recent methods \cite{weber2024nerfiller, signerf} employ inpainting pipelines over grids of image priors to achieve view-consistent edits.
Inspired by this strategy, we instead construct a tiled grid of sampled depth maps $D_{ref}$ as a unified reference input for our depth-conditioned multi-view generation pipeline.
Similar to Style-NeRF2NeRF \cite{fujiwara2024sn2n}, we incorporate an attention-sharing mechanism \cite{hertz2024style} to propagate consistent style across views.
Unlike the fully-shared attention approach proposed in prior work, however, our method anchors the generation process on tiled depth maps representing multiple views (arranged two-by-two in our experiments)  representing different views, enforcing structural alignment across diverse camera angles.
Empirically, we find that this reference-guided attention sharing further enhances appearance consistency, outperforming the closest prior method \cite{fujiwara2024sn2n}.

Following \cite{hertz2024style}, we replace the self-attention operation in the T2I diffusion model using the following formulas:

\begin{equation}
\texttt{Shared-Attn}(\hat{Q_t}, K_{rt}, V_{rt}) = \texttt{softmax} \left( \frac{\hat{Q_t} K_{rt}^T}{\sqrt{d_k}}V_{rt} \right)\\
\label{eq:shared-attn}
\end{equation}
\begin{equation}
K_{rt} = [K_{ref}, \hat{K}_t]^T, V_{rt} = [V_{ref}, V_t]^T
\label{eq:eq-shared-attention}
\end{equation}
where target queries $\hat{Q}_t$ and keys $\hat{K}_t$ are normalized via adaptive instance normalization \cite{huang2017arbitrary_adain} with reference queries $Q_{ref}$ and keys $K_{ref}$ of the tiled control reference $D_{ref}$ :
\begin{equation}
\hat{Q}_t = \texttt{AdaIN}(Q_t, Q_{ref}), \hat{K}_t = \texttt{AdaIN}(K_t, K_{ref})
\end{equation}
\begin{equation}
\texttt{AdaIN}(x, y) = \sigma(y) \Big( \frac{x-\mu(x)}{\sigma(x)} \Big) + \mu(y)
\end{equation}

\begin{figure*}[t]
  \centering
  \includegraphics[width=\textwidth]{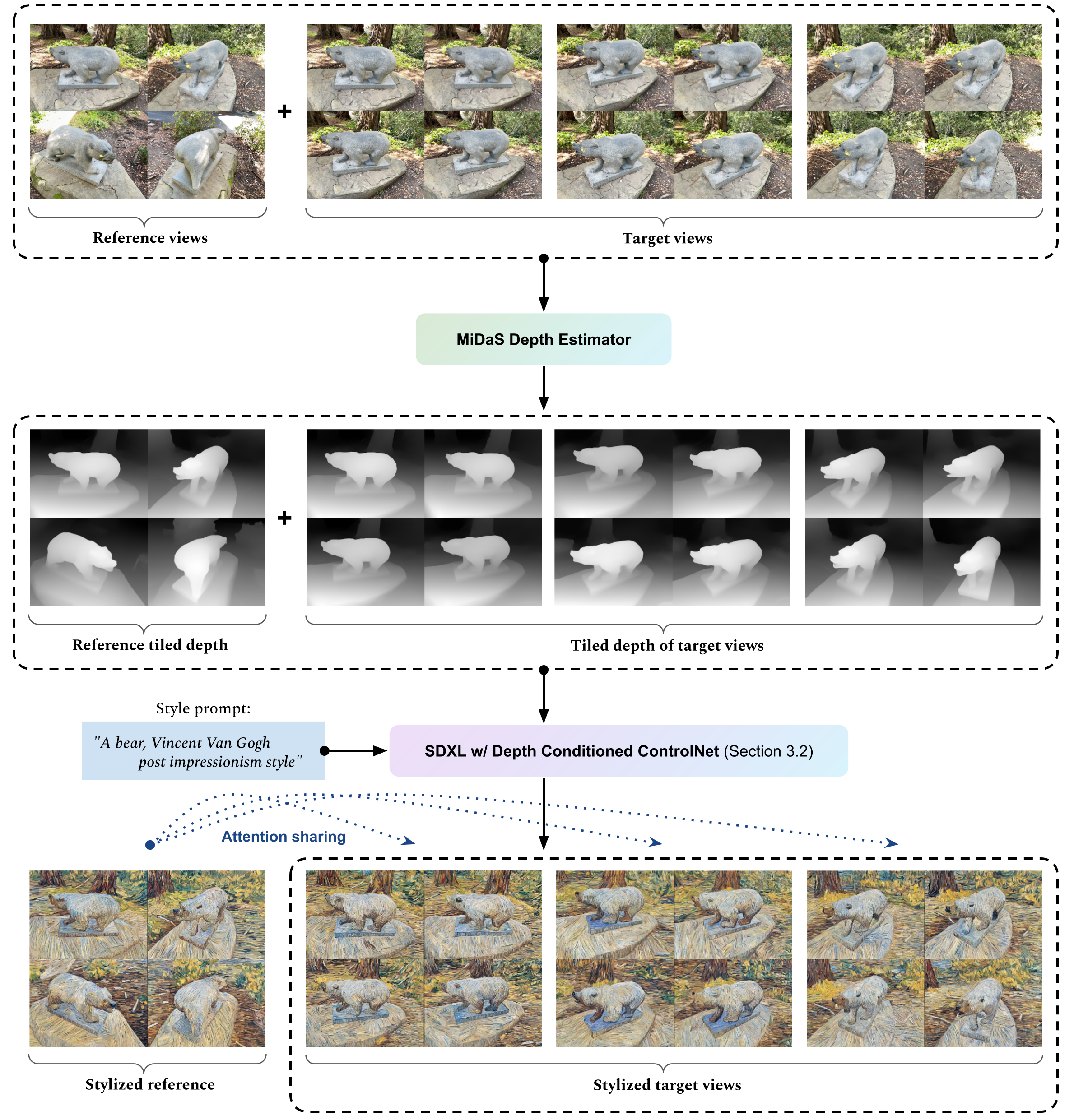}
  \caption{\textbf{Image-to-Image Multi-View Generation Pipeline}. We first obtain depth maps for both tiled representative views and the target views using an off-the-shelf depth prediction model. The tiled depth maps are then provided as conditioning input to a depth-guided ControlNet attached to a Stable Diffusion XL (SDXL) model. To ensure consistent appearance across viewpoints, the diffusion model incorporates an attention-sharing mechanism anchored on the reference tiled depth map, enabling coherent stylization of the target multi-view images.}
  \label{fig:mv_pipeline}
\end{figure*}

We present the complete procedures for the multi-view generation phase using 2$\times$2 image tiling in Algorithm \ref{alg:mv-algo}.
An illustration of our multi-view stylization pipeline is also shown in Figure \ref{fig:mv_pipeline}.

\begin{algorithm}[t]
\caption{Multi-View Editing w/ Tiled Depth Ref.}
\label{alg:mv-algo}
\KwIn{
    Set of source views $\mathcal{L}_{\text{in}} = \{I_1, I_2, \dots, I_n\}$;\\
    Text prompt $s$;\\
    Depth prediction model $M(I)$;\\
    SDXL w/ depth ControlNet $F(D_{\text{ref}}, D, s)$;
}
\KwOut{
    Set of stylized views $\mathcal{L}_{\text{out}} = \{I_1', I_2', \dots, I_n'\}$
}

$\mathcal{D}_{\text{in}} \leftarrow [M(I)\ \textbf{for}\ I \in \mathcal{L}_{\text{in}}]$\tcp*{Predict depth maps for input views}
$D_1, D_2, D_3, D_4 \leftarrow \text{sample\_ref\_depth}(\mathcal{D}_{\text{in}})$\tcp*{Select 4 diverse depth maps}

$D_{\text{ref}} \leftarrow \text{tile\_images}(D_1, D_2, D_3, D_4)$\tcp*{Create reference depth tile}

$\mathcal{L}_{\text{out}} \leftarrow \emptyset$\;

\While{$\mathcal{D}_{\text{in}}$ is not empty}{
    $D \leftarrow \text{pop\_four\_views}(\mathcal{D}_{\text{in}})$\tcp*{Handle edge case via duplication if needed}

    $D_{\text{tile}} \leftarrow \text{tile\_images}(D)$\;

    $I_{\text{tile}} \leftarrow F(D_{\text{ref}}, D_{\text{tile}}, s)$\tcp*{Stylize via attention anchoring on reference}

    $I_a', I_b', I_c', I_d' \leftarrow \text{untile\_images}(I_{\text{tile}})$\;

    $\mathcal{L}_{\text{out}}.\text{append}([I_a', I_b', I_c', I_d'])$\;
}

\Return $\mathcal{L}_{\text{out}}$
\end{algorithm}

\begin{figure*}[t]
  \centering
  \includegraphics[width=\textwidth]{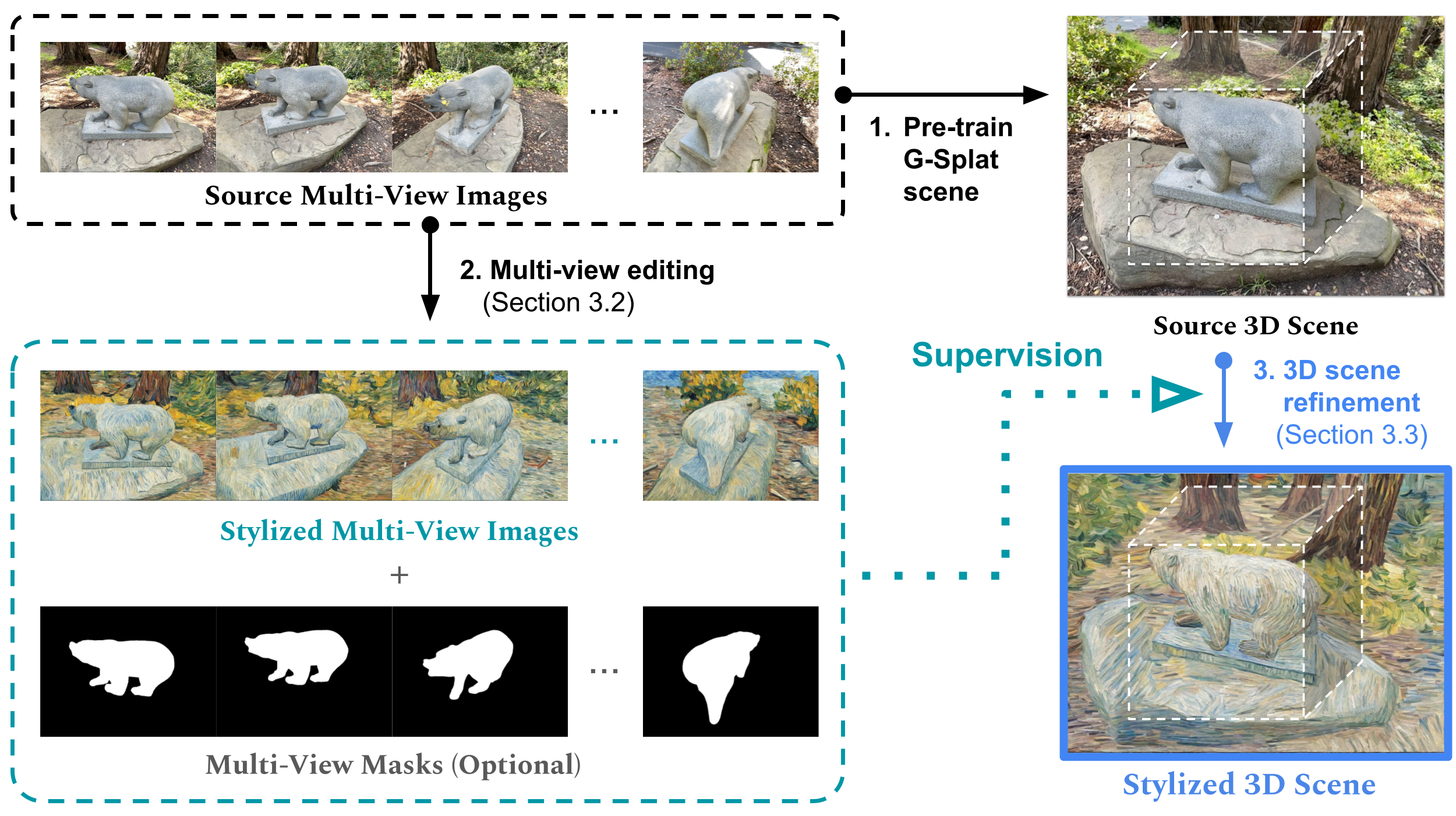}
  \caption{\textbf{Overall Multi-View Stylization Pipeline}. 1. We generate stylized multi-view images of the source scene using our custom diffusion pipeline anchored
on tiled depth maps. 2. Next, the source 3D scene is finetuned on the generated images. This refinement stage may take optional region masks for spatial control of the style transfer process (\ref{mr-iw-swd}).}
  \label{fig:overall_pipeline}
\end{figure*}

\subsection{3D Scene Refinement}
After generating stylized multi-view images, the goal is to fine-tune the underlying 3D representation to reflect the desired appearance. We now introduce several techniques to enhance this refinement process, including improved loss functions that encourage semantic alignment and promote more efficient convergence.

\subsubsection{Importance-Weighted Sliced Wasserstein Distance Loss}
Based on the assumption that a projection direction that gives a higher 1D Wasserstein distance value is more informative, we propose applying the energy-based SWD variant with importance sampling as studied by \cite{nguyen2023energy}.
Specifically, we find that extending the loss function in \ref{eq:sw} with importance weighting using Softmax function leads to enhanced training efficiency only with a minor implementation cost compared to the vanilla SWD loss:

\begin{align}
    \mathcal{L}_{IW\text{-}SWD} = \sum_{l=1}^{L} \sum_V w_V \mathcal{L}_{SW1D}(p^l_V, \hat{p}^l_V) \label{eq:iw-sw1} \\
    w_V = \frac{\exp(\mathcal{L}_{SW1D}(p^l_V, \hat{p}^l_V))}{ \sum_{V\prime} \exp(\mathcal{L}_{SW1D}(p^l_V\prime, \hat{p}^l_V\prime)) }
    \label{eq:iw-sw2}
\end{align}

We provide a comparative analysis in \ref{sec:loss-comparison} with an illustration of convergence performance in Figure \ref{fig:loss-comparison}.

\subsubsection{Multi-Region Importance-Weighted Sliced Wasserstein Distance Loss} \label{mr-iw-swd}
Additionally, we propose a region-based style transfer loss, which is similar to the concept introduced by a 2D neural texture synthesis method \cite{li2022sliced} and formulate the idea as follows. 

Given a $K$ categorical mask $b = (r_1, r_2, ..., r_{M_l}), r_m \in \{0,...,K\}$, we partition the projected scalar features $p^l_V \in \mathbb{R}^{M_l}$ into $K$ non-overlapping subsets $\{p^l_{V,k}\} ,(k=1,...,K)$ where $p^l_{V,k} =\{p^l_m | r_m=k \}$ is a vector containing elements of $p^l_V$ wherever $b$ has a value of $k$. Using this partitioned set of features, we extend the projected 1D Wasserstein loss in equation \eqref{eq:sw1d} as below:

\begin{multline}
    \mathcal{L}_{MR\text{-}SW1D}(p^l_{V, b}, \hat{p}^l_{V, b}) = \\ \sum^K_{k=1} \frac{1}{|p^l_{V,k}|} \lVert \texttt{sort}(p^l_{V,k}) - \texttt{sort}(\hat{p}^l_{V,k}) \rVert^2    
\label{eq:masked-sw1d}
\end{multline}

\begin{figure*}[t]
  \centering
  \includegraphics[width=\textwidth]{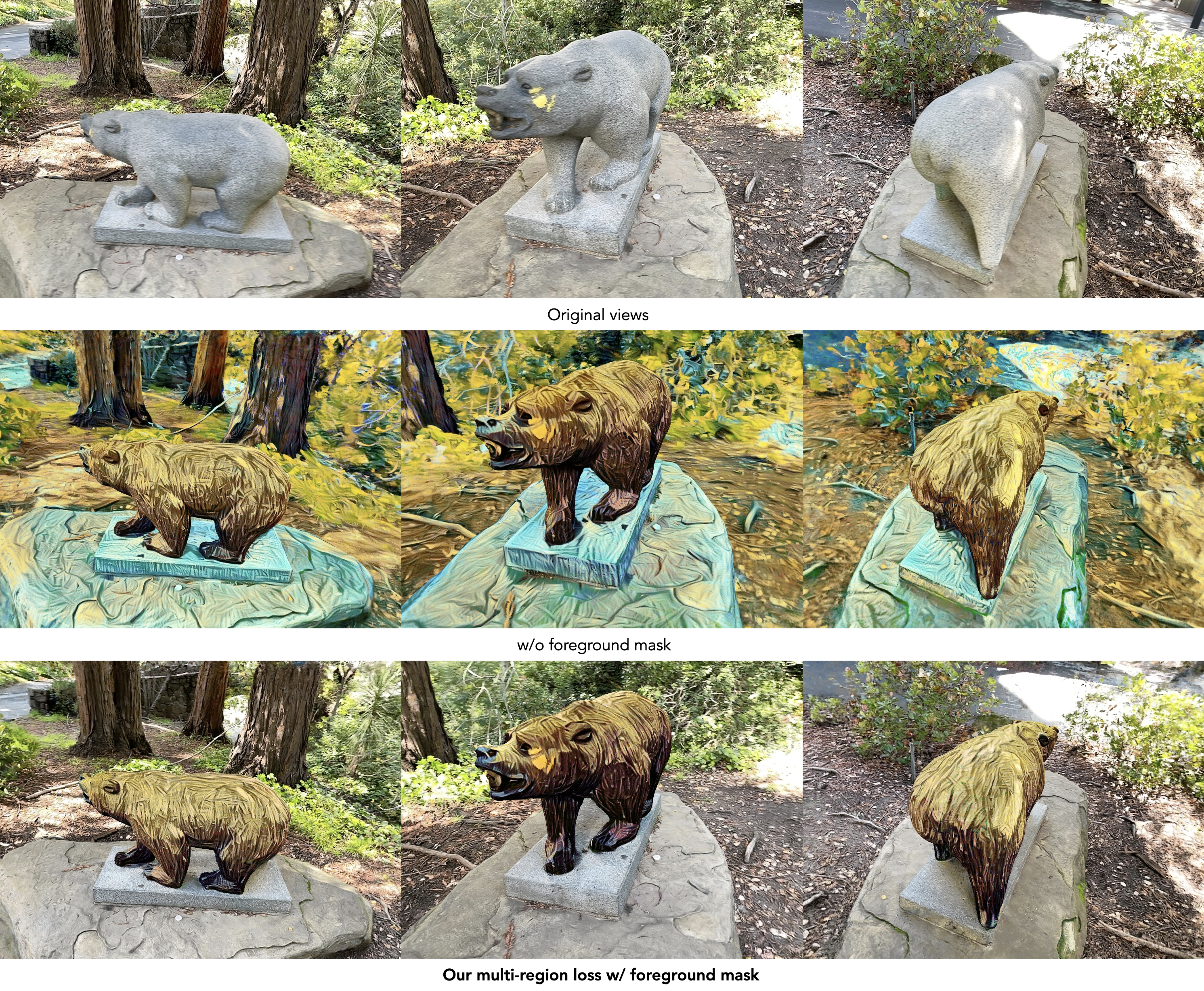}
  \caption{\textbf{Multi-region loss application example with a binary mask.} A binary mask is used to apply stylization only to the foreground object—in this case, the bear statue—while preserving the original appearance of the background. This demonstrates the ability of our method to perform selective, region-specific 3D stylization.}
  \label{fig:binmask}
\end{figure*}

\begin{figure*}[t]
  \centering
  \includegraphics[width=\textwidth]{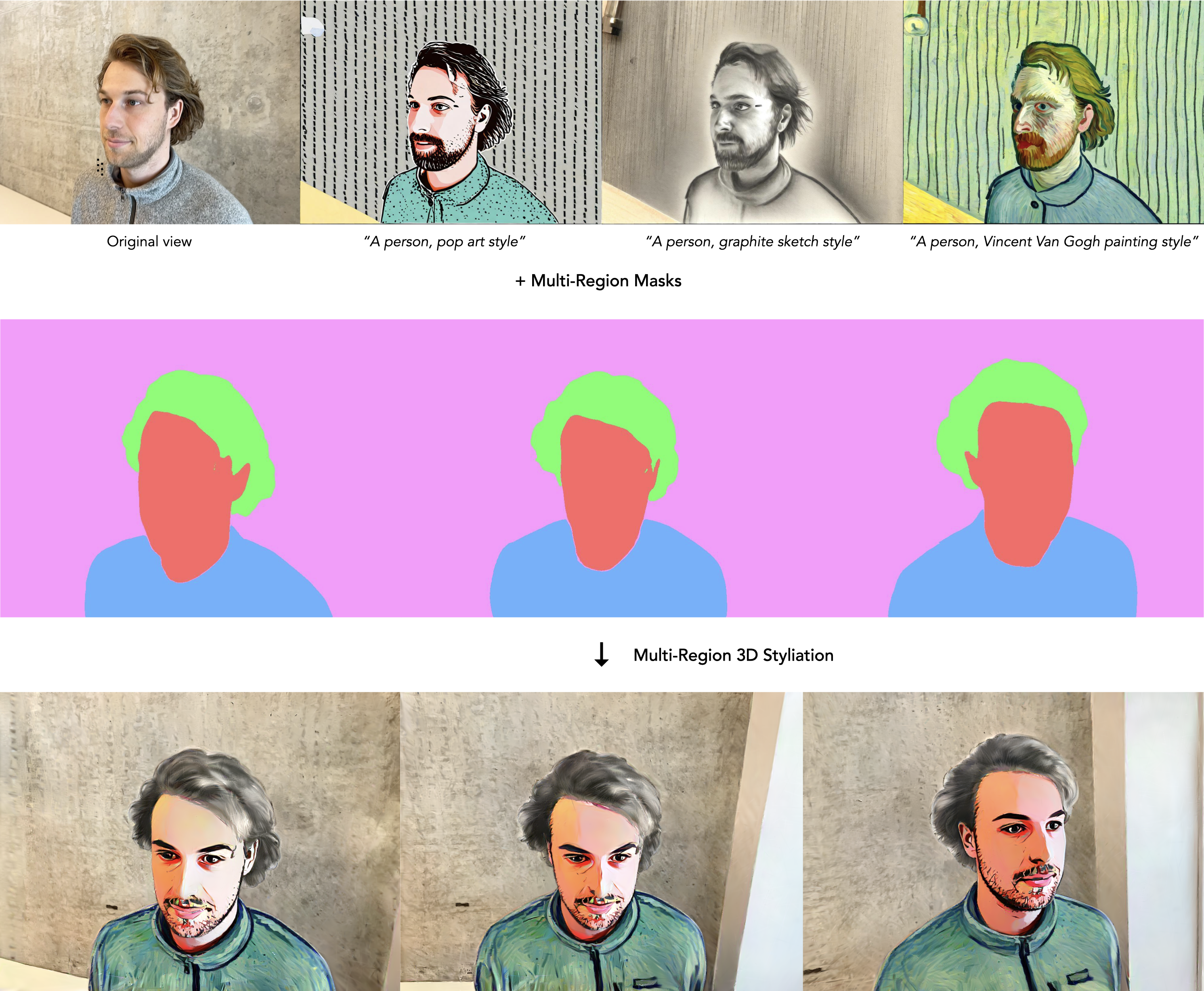}
  \caption{\textbf{Example of 3D scene stylization using our multi-region loss.} Three distinct styles are applied to different semantic regions based on the segmentation masks, enabling spatially controlled stylization. In this example, the background region is excluded from stylizatixon and remains unchanged, demonstrating the ability to selectively preserve parts of the original scene.}
  \label{fig:multiregion}
\end{figure*}

Using this partitioned set, we extend the 1D projected Wasserstein loss in Equation (4) to a region-aware formulation that computes the loss separately for each region.
This partitioning enables spatial control by aligning feature distributions independently within each semantic region.
When a mask is provided, the standard loss $\mathcal{L}_{SW1D}$ in Equations \eqref{eq:iw-sw1} and \eqref{eq:iw-sw2} can be replaced with the multi-region variant $\mathcal{L}_{MR\text{-}SW1D}$ to activate this feature.
Typical use cases include applying a binary mask (e.g. $K=1$) to perform style transfer separately on foreground and background regions, or stylizing only the foreground while leaving the background unchanged.
Importantly, our use of Gaussian splatting enables computing this loss over full images, unlike NeRF-based stylization methods that require patch-based sampling due to memory constraints.
This feature contributes to maintaining style consistency within specific semantic regions.
In summary, our multi-region loss function supports (1) semantically consistent style transfer, (2) applying style selectively to targeted areas of the scene (Figure \ref{fig:binmask}), and (3) blending different styles across regions (Figure \ref{fig:multiregion}).

\subsubsection{Content Loss}
To maintain the original structure of the scene, we further incorporate a content preservation loss \cite{li2022sliced} by calculating the mean-squared error of VGG19 activations:
\begin{equation}
\mathcal{L}_{content} = \lambda \sum_l \| F^l-\hat{F}^l \|^2
\label{eq:content-loss}
\end{equation}
We set the weight to $\lambda = 0.1$ in all our experiments.

\section{Experiments}
\subsection{Experimental Setup}
We evaluate our method on multiple scenes selected from Instruct-NeRF2NeRF dataset \cite{instructnerf2023} and Mip-NeRF360 dataset \cite{barron2022mip}, which provide diverse and high-quality 3D scenes. Our 2D multi-view stylization pipeline is built on top of an SDXL-based style-aligned diffusion framework \cite{hertz2024style}, enabling text-driven appearance editing. 
To obtain the depth maps for the tiled control image, we use MiDAS \cite{ranftl2020towards}.
For 3D scene refinement, we clone the official implementation of 2D Gaussian Splatting (2DGS) \cite{huang20242d} as our representation backbone.
We use SAM2 \cite{ravi2024sam2} to obtain segmentation masks for our multi-region loss in the stylization process.
All baseline methods are evaluated using the publicly released code provided by their respective authors. We train all scenes on a single NVIDIA A100 GPU, running for 500–1000 iterations, which takes approximately 10-15 minutes depending on the image resolution and number of views in the scene.

\subsection{Text-Driven 3D Stylization}
We compare our method against two state-of-the-art approaches: Style-NeRF2NeRF~\cite{fujiwara2024sn2n} and DGE~\cite{chen2024dge}. Both methods follow a similar approach, where the source 3D scene is directly edited using a set of consistently stylized multi-view images generated from text prompts. While sharing this overall structure, our method introduces key improvements in the multi-view generation pipeline and 3D refinement stage, leading to improved visual fidelity and semantic alignment with the input prompts. Qualitative comparisons and quantitative results are summarized in Table~\ref{tab:quantitative-results} and Figure~\ref{fig:comparison}.

\subsubsection{Qualitative Results}
\begin{figure*}[t]
  \centering
  \includegraphics[width=0.8\textwidth]{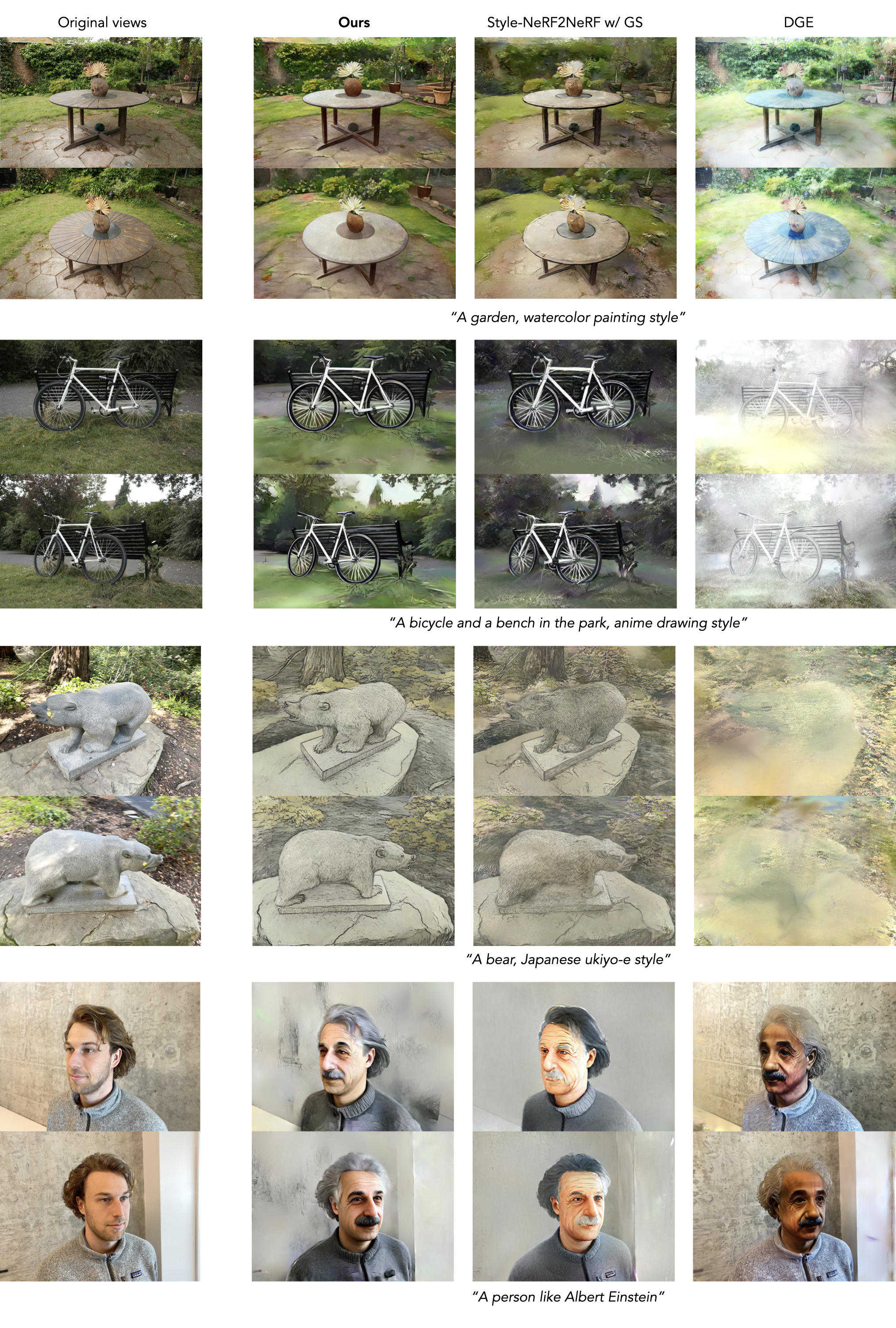}
  \caption{\textbf{Method Comparison.} We compare our method against Style-NeRF2NeRF~\cite{fujiwara2024sn2n} and DGE~\cite{chen2024dge}. As shown, our approach produces clearer and more visually artistic results that are faithful to the given style prompts, while exhibiting fewer artifacts. For a fair comparison, we replace the underlying NeRF representation in the Style-NeRF2NeRF baseline with the same 2D Gaussian Splatting (2DGS)~\cite{huang20242d} used in our method.}
  \label{fig:comparison}
\end{figure*}

We present qualitative comparisons of our 3D stylization results in Figure~\ref{fig:comparison}, alongside results from prior methods.
Since we use the same diffusion backbone (SDXL~\cite{podell2023sdxl}) as Style-NeRF2NeRF~\cite{fujiwara2024sn2n} and fix the random seed across experiments, the overall stylization atmosphere appears similar.
However, our method yields results that are more faithful to the style prompts, with sharper details and fewer visual artifacts. 
While DGE demonstrates strong performance in portrait-style edits, it occasionally fails to stylize the full scene coherently and often produces over-saturated colors.
For a more comprehensive view, we encourage readers to watch the demo video.

\subsubsection{Quantitative Results}
We quantitatively evaluate the performance of our method using standard metrics following prior work \cite{gal2022stylegan}, as summarized in Table \ref{tab:quantitative-results}.
Specifically, we report CLIP similarity, which measures image-text alignment between the source and stylized images in CLIP’s embedding space.
To assess view consistency, we also compute warping errors across sequential frames using optical flow estimated by RAFT~\cite{teed2020raft}, which captures temporal coherence from a geometric perspective.
We calculate the average mean squared error (MSE) between the forward-warped $i-1$-th frame and the $i$-th frame.
As shown in Table~\ref{tab:quantitative-results}, our method achieves competitive performance across all metrics when compared to baseline approaches.
That said, we acknowledge that stylization is inherently subjective, and user preferences may vary depending on specific goals.
To complement the quantitative evaluation, we conducted a user study involving 58 participants.
Each participant was shown 12 randomly ordered scene–style prompt pairs (4 scenes × 3 styles), and asked to choose the most visually appealing result for each.
In total, 696 responses were collected (58 participants × 12 questions), and the aggregated preferences are also reported in Table~\ref{tab:quantitative-results}.

\begin{table}[h]
    \centering
    \renewcommand{\arraystretch}{1.5} % 1.0 is default
    \begin{tabular}{c|ccc}
        \hline
        Method & \makecell{CLIP \\ Similarity $\uparrow$}  & \makecell{Warp \\ Error $\downarrow$} & \makecell{User \\ Pref. $\uparrow$} \\
        \hline
        \makecell{Style-N2N \\ w/ GS \cite{fujiwara2024sn2n}} & 0.142 & 0.054 & 19.9\% \\
        \hline
        \makecell{DGE \\ \cite{chen2024dge}} & 0.184 & 0.072 & 22.0\% \\
        \hline
        \textbf{Ours} & \textbf{0.213} & \textbf{0.050} & \textbf{58.8}\% \\
        \hline
    \end{tabular}
    \caption{\textbf{Method Comparisons.} We report quantitative values using metrics following common literature. For the Style-NeRF2NeRF baseline \cite{fujiwara2024sn2n}, we replace their backbone 3D representation with the same 2D Gaussian splatting \cite{huang20242d} as ours to make a fair comparison.}
    \label{tab:quantitative-results}
\end{table}

\subsection{Ablations}
To validate the effectiveness of our proposed components, we perform ablation experiments by comparing our full method with several simplified variants.
The first variant replaces our tiled depth map reference with a single depth map and employs a fully-shared attention diffusion pipeline, as proposed in prior work~\cite{fujiwara2024sn2n}.
The second variant turns off the multi-region functionality by omitting the input masks, allowing us to assess the contribution of our multi-region SWD loss.
We also verify that skipping content loss substantially increases the Warp Error, proving that content loss contributes to maintaining 3D view consistency.

As illustrated in Figure \ref{fig:ablation}, the absence of our multi-region loss results in visible color bleeding; for example, the color or style of a particular object may unintentionally appear in unrelated parts of the scene.
Furthermore, when compared to the single-depth variant, our proposed multi-view generation pipeline produces stylized results that are sharper and exhibit fewer artifacts, demonstrating the benefit of view-aware conditioning and attention anchoring.
We also include quantitative evaluation results in Table \ref{tab:ablation-results} for reference.

\begin{table}[h]
    \centering
    \renewcommand{\arraystretch}{1.5} % 1.0 is default
    \begin{tabular}{c|cc}
        \hline
        Variant & \makecell{CLIP \\ Similarity $\uparrow$}  & \makecell{Warp \\ Error $\downarrow$}  \\
        \hline
        \makecell{w/o tiled ref. MV pipeline} & 0.178 & 0.061 \\
        \hline
        \makecell{w/o multi-region loss} & 0.201 & 0.058 \\
        \hline
        \makecell{w/o content loss} & 0.210 & 0.064 \\
        \hline
        \textbf{Ours} & \textbf{0.213} & \textbf{0.050} \\
        \hline
    \end{tabular}
    \caption{\textbf{Ablation Study.} We report evaluation metrics for different variants of our method to assess the impact of each proposed component.}
    \label{tab:ablation-results}
\end{table}

\subsection{Importance-Weighted SWD v.s. Vanilla SWD}
\label{sec:loss-comparison}
\begin{figure}[ht]
  \centering
  \includegraphics[width=\columnwidth]{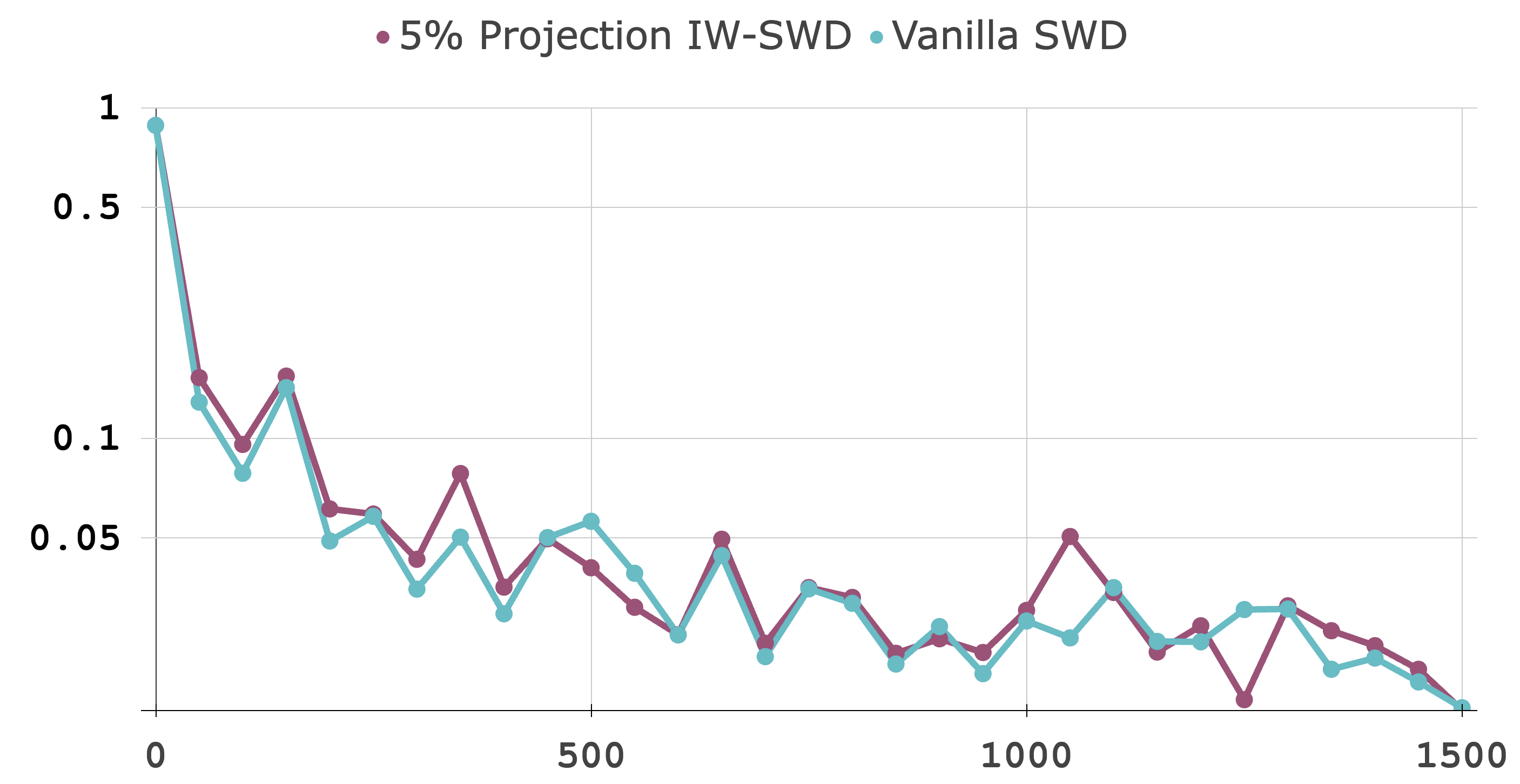}
  \caption{\textbf{Comparison of loss convergence between vanilla SWD and our proposed IW-SWD.} We plot the SWD loss values over 1,500 training iterations for both vanilla SWD and our importance-weighted SWD (IW-SWD), which uses only 5\% of the projection samples. Despite the reduced number of projections, IW-SWD achieves comparable convergence behavior, demonstrating its efficiency in guiding 3D stylization with significantly fewer computations.}
  \label{fig:loss-comparison}
\end{figure}

Finally, we evaluate the performance of the importance-weighted SWD, which incorporates a Softmax-based weighting to prioritize informative projections.
To illustrate the convergence behavior, we plot the average SWD loss values across multiple scenes at every 50 training iterations, up to a total of 1,500 iterations. As shown in Figure~\ref{fig:loss-comparison}, IW-SWD achieves comparable convergence performance to the vanilla SWD, despite using only 5\% of the slicing projections. This result confirms that IW-SWD enables significantly more efficient stylization, maintaining quality while greatly reducing the number of required projection samples.

\begin{table}[h]
    \centering
    \renewcommand{\arraystretch}{1.5} % 1.0 is default
    \begin{tabular}{c|cc}
        \hline
        Method & Runtime(msec) $\downarrow$  & TeraFLOPS $\downarrow$  \\
        \hline
        Uniformly Weighted & 421 & 1.174 \\
        \hline
        \textbf{IW w/ 5\% Projection} & \textbf{118} & \textbf{1.035} \\
        \hline
    \end{tabular}
    \caption{\textbf{IW-SWD vs Vanilla SWD.} We compare the performance of uniformly weighted SWD against importance-weighted SWD with only 5\% projections by measuring the average runtime and TeraFLOPS of the loss computation within a single training iteration. The only differences are fewer projections and the Softmax weighting of 1D-SWD values for the sampled projection directions, while keeping the same VGG19 backbone.}
    \label{tab:swd-comparison-results}
\end{table}

\iffalse
\begin{figure*}[t]
  \centering
  \includegraphics[width=\linewidth]{figures/ablation.png}
  \caption{Ablation results for 3D style transfer using the prompt \textit{"A painting of a blue bear."} Without our multi-region loss, stylization exhibits color bleeding, with blue regions spilling outside the bear. Additionally, our multi-view generation pipeline produces sharper results with fewer artifacts, demonstrating improved consistency and fidelity in the stylized 3D scene.
}
  \label{fig:ablation}
\end{figure*}
\fi

\section{Future Work and Limitations}
Main limitation of our method is that it relies on depth conditioning, which means it cannot significantly modify the underlying geometry or shape of the scene.
In the future, this could be addressed by incorporating generative models that are more geometry-aware.
Another interesting direction is to extend our approach to dynamic scenes, enabling stylization that remains consistent over time.
\section{Conclusion}
We presented a 3D scene stylization framework based on Gaussian Splatting, structured as a two-step process: text-driven multi-view stylization followed by 3D source scene refinement. Our training-free diffusion pipeline, conditioned on a grid of sampled depth maps, enables style-consistent multi-view generation, while the proposed multi-region IW-SWD loss facilitates semantically robust and efficient fine-tuning of the 3D scene. Notably, the multi-region formulation also lends itself to more flexible applications, such as partial scene stylization and multi-style composition. Our method offers an intuitive and competitive solution for 3D scene stylization. 

\section*{Acknowledgements}
This work was partially supported by JST Moonshot R\&D Grant Number JPMJPS2011, CREST Grant Number JPMJCR2015 and Basic Research Grant (Super AI) of Institute for AI and Beyond of the University of Tokyo.

% Figures only pages
\clearpage

\clearpage

\begin{figure*}[t]
  \centering
  \includegraphics[width=\linewidth]{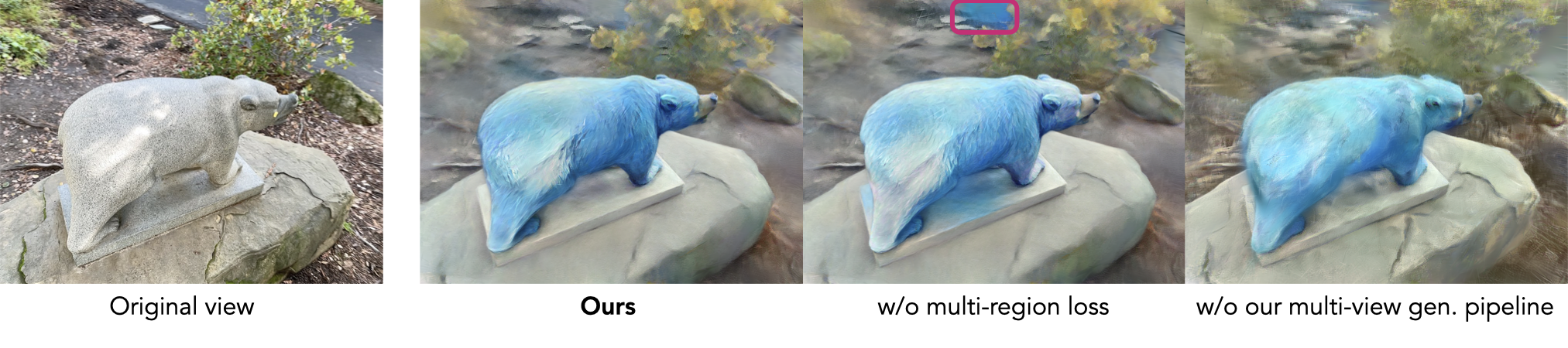}
  \caption{Ablation results for 3D style transfer using the prompt \textit{"A painting of a blue bear."} Without our multi-region loss, stylization exhibits color bleeding, with blue regions spilling outside the bear. Additionally, our multi-view generation pipeline produces sharper results with fewer artifacts, demonstrating improved consistency and fidelity in the stylized 3D scene.
}
  \label{fig:ablation}
\end{figure*}

\clearpage   

\clearpage   

% bibtex
\bibliographystyle{eg-alpha-doi} 
\bibliography{egbibsample}

\clearpage
\renewcommand{\thesection}{\Alph{section}}
\setcounter{section}{0}

% Content is the same as suuplementary.tex

\twocolumn[
\begin{center}
  \vspace{4em}
  {\LARGE \bfseries Supplementary Material \par}
  \vspace{4em}
\end{center}
]

% Content
\section{Additional Implementation Details}

\subsection{2D Stylization Pipeline Settings}
We implemented our 2D multi-view stylization pipeline by building on the code from \cite{hertz2024style}, which uses Stable Diffusion XL \cite{podell2023sdxl} as a backbone. We attach depth-conditioning ControlNet \cite{zhang2023adding} released by Diffusers from Huggingface. For view generation, we set the ControlNet conditioning scale to $1.0$ with $50$ diffusion steps. We also find that a classifier guidance scale within the range $[4.5,13.5]$ usually works well, depending on the stylization objective. We encourage users to test our released code for further details.

\subsection{Loss Functions}
For pre-training of the source scene, we follow the same setup as \cite{huang20242d}.
During the scene stylization stage, we disable the L1 photometric loss and add our proposed IWD-SWD loss, which calculates the (importance-weighted) sliced Wasserstein distance between VGG19 feature maps of the rendered image and the corresponding ground truth image. We use the first 12 layers of VGG19 to extract the feature maps. The number of projections are 3, 6, 13, and 26 ($5\%$ of 64, 128, 256, and 512) for layer $l=1,2$, $l=3,4$, $l=5,6,7,8$, and $l=9,10,11,12$ respectively. The code for IW-SWD is based on the implementation released by \cite{heitz2021sliced}.

\section{Importance-Weighted Sliced Wasserstein Distance}
In this section, we leave supplementary notes regarding the formulation of the Importance-Weighted Sliced Wasserstein Distance (IW-SWD).
For readers interested in more theoretical details, we strongly recommend referring to the original paper \cite{nguyen2023energy}.

\subsection{Sliced Wasserstein}
We denote $\mathcal{P}(\chi)$ as the set of probability measures over a set $\chi$.
Given two probability measures, $\mu, \nu \in \mathcal{P}(\mathbb{R}^d)$, the sliced Wasserstein distance is defined as:
\begin{equation}
    SW(\mu, \nu) = \big( \mathbb{E}_{V \sim \mathcal{U}(\mathbb{S} ^{d-1})} [ W^p_p(V\sharp\mu, V\sharp\nu)] \big)^{\frac{1}{p}}
\label{eq:sw}
\end{equation}
where $V$ is a projection vector following a uniform distribution over a unit hyper-sphere $\mathcal{U}(\mathbb{S}^{d-1})$ and $V\sharp\mu, V\sharp\nu$ are push-forward measures of $\mu, \nu$ respectively. The intractable expectation in equation \ref{eq:sw} is approximated via the Monte Carlo scheme:
\begin{equation}
    \widehat{SW}(\mu, \nu) =\Big( \frac{1}{K} \sum^K_{k=1}  W^p_p(V_k\sharp\mu, V_k\sharp\nu) \Big)^{\frac{1}{p}}
\label{eq:sw-montecarlo}
\end{equation}
where ${V_k} \sim \mathcal{U}(\mathbb{S}^{d-1})$ are $K$ projection vectors sampled from a unit hyper-sphere. Since $\mu, \nu$ are discrete measures representing image features, time complexity is $\mathcal{O}(Kn\log(n))$ when $\mu, \nu$ have at most $n$ supports. $\widehat{SW}$ is proven to be an unbiased estimation of $SW$ with the property $\widehat{SW}(\mu, \nu; K) \rightarrow SW(\mu, \nu)$ for $K \rightarrow \infty$.

\subsection{Energy-Based Sliced Wasserstein}
Although $\widehat{SW}(\mu, \nu)$ serves as a useful term for general machine learning tasks, some slicing directions from uniform sampling may not be as informative towards the true expectation value $SW(\mu, \nu)$.
To address this issue, \cite{nguyen2023energy} introduced the Energy-Based Sliced Wasserstein (EBSW) distance as an optimization-free variant that exploits the value of the projected Wasserstein distance between $\mu$ and $\nu$.

Based on the fundamental assumption: \textit{"A higher value of projected Wasserstein distance, a better projecting direction"}, they define an energy function $f:[0, \infty) \rightarrow \Xi \in (0, \infty)$ and an energy-based slicing distribution $\sigma_{\mu,\nu}(V)$ supported on $\mathbb{S}^{d-1}$ as the following:
\begin{equation}
    \sigma_{\mu, \nu}(V) \propto f(W^p_p(V\sharp\mu, V\sharp\nu)) = \frac{f(W^p_p(V\sharp\mu, V\sharp\nu))}{ \int_{\mathbb{S}^{d-1}} f(W^p_p(V\sharp\mu, V\sharp\nu)) dV}
\label{eq:ebswd-dist}
\end{equation}
where monotonically increasing energy functions such as the exponential function $f(x) = e^x$ are recommended from a practical standpoint.

Using the energy-based slicing distribution given in the definition \ref{eq:ebswd-dist}, EBSW distance is defined as below:
\begin{equation}
    EBSW_p(\mu, \nu; f) = \big( \mathbb{E}_{V \sim \sigma_(\mu, \nu)(V;f)} [ W^p_p(V\sharp\mu, V\sharp\nu) ] \big)^{\frac{1}{p}}
\label{eq:ebswd-def}
\end{equation}
which is also the upper bound of SW for any $p \geq 1$ and an increasing energy function $f$.
\begin{equation}
    SW_p(\mu, \nu) \leq EBSW_p(\mu, \nu, f)
\label{eq:ebswd-upper}
\end{equation}

The authors also propose several sampling schemes for EBSW approximation. Among them, we use the Monte Carlo sampling based on importance sampling \cite{kloek1978bayesian}. The EBSW distance using a sampling proposal distribution $\sigma_0(V) \in \mathcal{P}(\mathbb{S}^{d-1})$ can be reformulated as the following:
\begin{equation}
    EBSW_p(\mu, \nu; f) = \Big( \frac{\mathbb{E}_{V \sim \sigma_0(V)} [ W^p_p(V\sharp\mu, V\sharp\nu) w_{\mu, \nu, \sigma_0, f, p}(V) ]}{ \mathbb{E}_{V \sim \sigma_0(V)} [w_{\mu, \nu, \sigma_0, f, p}(V)] } \Big)^{\frac{1}{p}}
\label{eq:ebswd-re}
\end{equation}
\begin{equation}
    w_{\mu, \nu, \sigma_0, f, p}(V) = \frac{f(W^p_p(V\sharp\mu, V\sharp\nu))}{\sigma_0(V)}
\label{eq:ebswd-weight}
\end{equation}
Using \textit{i.i.d.} samples $V_1, ..., V_K \in \sigma_0(V)$, the importance sampling estimator IS-EBSW is formulated as:
\begin{equation}
    \widehat{IS\text{-}EBSW}_p(\mu, \nu; f, K) = \Big( \sum^K_{k=1} \hat{w}_{\mu, \nu, \sigma_0, f, p}(V_k) \cdot W^p_p(V_k\sharp\mu, V_k\sharp\nu)  \Big)^{\frac{1}{p}}
\label{eq:ebswd-is}
\end{equation}
\begin{equation}
    \hat{w}_{\mu, \nu, \sigma_0, f, p}(V_k) = \frac{w_{\mu, \nu, \sigma_0, f, p}(V_k)}{ \sum^K_{k'=1} w_{\mu, \nu, \sigma_0, f, p}(V_{k'}) }
\label{eq:ebswd-weight}
\end{equation}

Specifically, IW-SWD corresponds to this IS-EBSW case presented in the paper \cite{nguyen2023energy}, where we adopt the exponential energy function $f(x) = e^x$ with a uniform proposal distribution $\sigma_0(V) = \mathcal{U}(\mathbb{S}^{d-1})$, which is equivalent to weighting the SWD values using the Softmax function.

\section{Additional Comparisons}
\begin{figure*}[t]
  \centering
  \includegraphics[width=0.8\textwidth]{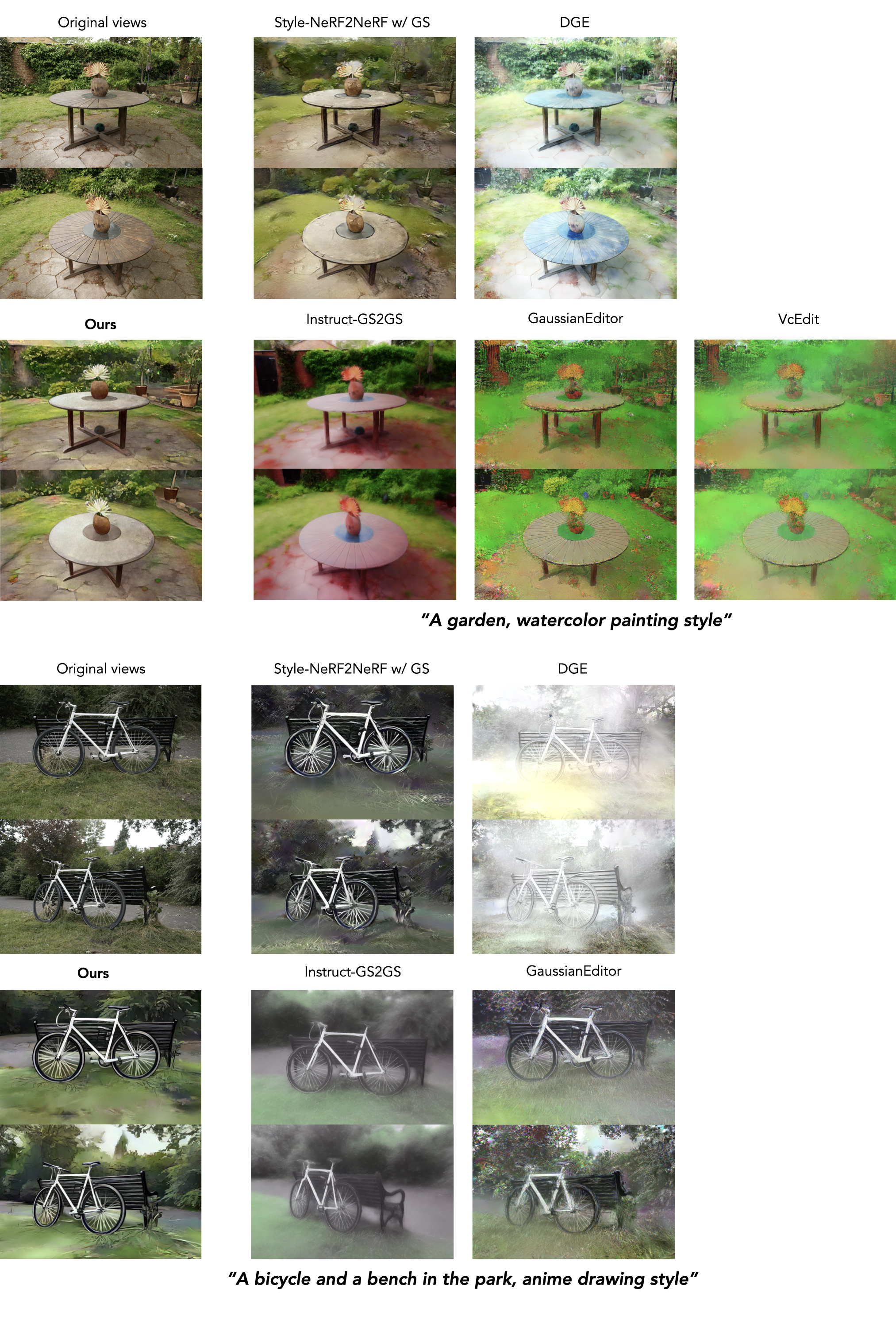}
  \caption{\textbf{Additional Method Comparison 1.} Our method exhibits artistically superior results without blurry artifacts. While VcEdit \cite{wang2024view} can perform convincing editing in some cases, we found it to be memory-intensive (~80GB of GPU memory with full training views) and relatively difficult for large-scale 360-degree scenes to converge when using holistic style transfer prompts.}
  \label{fig:comparison-supp1}
\end{figure*}

\begin{figure*}[t]
  \centering
  \includegraphics[width=0.8\textwidth]{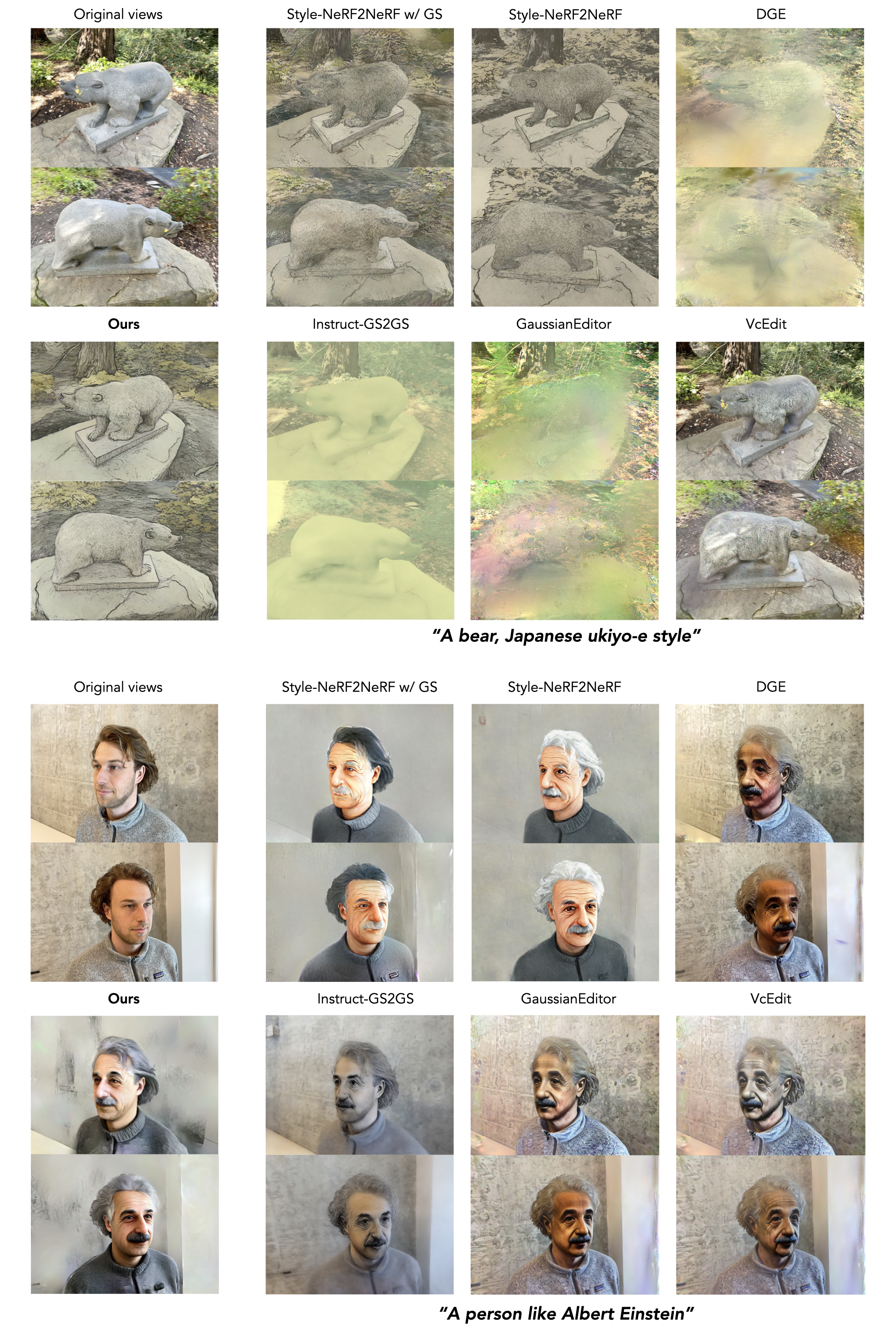}
  \caption{\textbf{Additional Method Comparison 2.} Our method can perform scene stylization with competitive performance, while other techniques may result in over-saturated colors or insufficient edits.}
  \label{fig:comparison-supp2}
\end{figure*}

In figure \ref{fig:comparison-supp1} and \ref{fig:comparison-supp2}, we include additional visual comparisons against Instruct-GS2GS \cite{igs2gs}, GaussianEditor \cite{chen2024gaussianeditor}, VcEdit \cite{wang2024view}, and Style-NeRF2NeRF (w/ original NeRF representation as baseline) \cite{fujiwara2024sn2n} for reference. The results are produced using the default settings for each method. Therefore, please note that tuning parameters such as the guidance scale of the underlying Instruct-Pix2Pix model \cite{brooks2023instructpix2pix} may result in different outcomes. We recommend that users test different methods according to their varying goals and requirements.

%-------------------------------------------------------------------------

% Figures only pages
\clearpage

\clearpage   

% Content End
\end{document}